\documentclass[preprint,showpacs,showkeys,aps]{revtex4}
\usepackage{graphicx}
\usepackage{dcolumn}
\usepackage{bm}
\begin{document}
\newcommand{\ti}[1]{\mbox{\tiny{#1}}}
\newcommand{\im}{\mathop{\mathrm{Im}}}
\def\be{\begin{equation}}
\def\ee{\end{equation}}
\def\bea{\begin{eqnarray}}
\def\eea{\end{eqnarray}}

\title{Quadrupolar gravitational fields described by the $q-$metric}

\author{Hernando Quevedo$^{1,2}$, Saken Toktarbay$^3$ and Aimuratov Yerlan$^3$ }
\affiliation{
$^1$Instituto de Ciencias Nucleares, Universidad Nacional Aut\'onoma de M\'exico, AP 70543, M\'exico, DF 04510, Mexico\\
$^2$Dipartimento di Fisica and ICRA, Universit\`a di Roma "La Sapienza", I-00185 Roma, Italy\\
$^3$Department of Physics, Al-Farabi Kazakh National University, 050040 Almaty, Kazakhstan}
\email{quevedo@nucleares.unam.mx,saken.yan@yandex.com}

\date{\today}

\begin{abstract}
We investigate the Zipoy-Voorhees metric ($q-$metric) as the simplest static, axially symmetric solution of Einstein's vacuum field equations 
that possesses as independent parameters the mass and the quadrupole moment. In accordance with the black holes uniqueness theorems,
the presence of the quadrupole completely changes the geometric properties of the corresponding spacetime that turns out to contain naked
singularities for all possible values of the quadrupole parameter. The naked singularities, however, can be covered by interior solutions 
that correspond to perfect fluid sources with no specific equations of state. We conclude that the $q-$metric can be used to describe the 
entire spacetime generated by static deformed compact objects.

\end{abstract} 

\maketitle

\section{Introduction}
The Zipoy--Voorhees metric  \cite{zip66,voor70} was discovered more than forty years ago as a particular exact solution 
of Einstein's vacuum field equations that belongs to the Weyl class \cite{solutions} of 
vacuum solutions. In this work, we will refer to the Zipoy-Voorhees solution  as to the $q-$metric  for a reason 
that will be explained below. Since its discovery, many works have been devoted to the investigation of its geometric and physical properties. 
In 
particular, it has been established that it describes an asymptotically flat spacetime, it possesses two commuting, hypersurface
orthogonal Killing vector fields that imply that the spacetime is static and axially symmetric,  it contains the Schwarzschild metric 
as a special case that turns out to be the only one with a true curvature singularity surrounded by an event horizon \cite{zv1,zv2,zv3,zv4,zv5}. 

In a recent work \cite{quev11}, it was proposed to interpret the $q-$metric as describing the gravitational field of a distribution of mass whose non-spherically 
symmetric shape is represented by an independent quadrupole parameter. Moreover, the curvature singularities turn out to be localized
 inside a region situated very close to the origin of coordinates. Consequently, this metric can be used to describe the exterior gravitational field of
 deformed distributions of mass in which the quadrupole moment is the main parameter that describes the deformation. Tue question arises whether it is possible 
 to find an interior metric that can be matched to the exterior one in such a way that the entire spacetime is described. 
 To this end, it is usually assumed that the interior mass distribution can be described by means of a perfect fluid with two physical 
 parameters, namely, energy density and pressure. The energy-momentum tensor of the perfect fluid is then used in the Einstein equations as the source of
 the gravitational field.  
 It turns out that the system of the corresponding differential equations cannot be solved, because the number of equations is less than the number of unknown functions. This problem is usually solved by imposing equations of state that relate the pressure and density of the fluid. In this work, however, we
 will explore a different approach that was first proposed by Synge \cite{synge}. To apply this method, one first uses general physical considerations
 to postulate the form of the interior metric and then one evaluates the energy-momentum tensor of the source by using Einstein's equations. In this manner, any 
 interior metric can be considered as an exact solution of the Einstein equations for some energy-momentum tensor. However, the main point of the procedure
 is to impose physical conditions on the resulting matter source so that it corresponds to a physical reasonable configuration. In general, one can impose
 the energy conditions, the matching conditions with the exterior metric, and conditions on the behavior of the metric functions near the center 
 of the source and on the boundary with the exterior field.

 This work is organized as follows. In Sec. \ref{sec:zv}, we review the main properties of the Zipoy-Voorhees transformation in different coordinate systems. 
 In Sec. \ref{sec:qmet}, we consider the $q-$metric as describing the exterior gravitational field of a deformed source with mass and quadrupole 
 moment. In Sec. \ref{sec:int}, we propose a particular interior solution and derive the corresponding energy-momentum tensor by using the Einstein
 equations. Finally, Sec. \ref{sec:con} contains discussions of our results and suggestions for further research.

 \section{The Zipoy--Voorhees transformation}
 \label{sec:zv}
 
Zipoy \cite{zip66} and Voorhees \cite{voor70} investigated static, axisymmetric vacuum solutions of Einstein's equations and  
found a simple transformation which allows to generate new solutions from a known solution. To illustrate the idea of the transformation,
we use the general line element for static, axisymmetric vacuum gravitational fields in prolate spheroidal coordinates
$(t,x,y,\varphi)$:
\be
 ds^2 =e^{2\psi} dt ^2 - \sigma^2 e^{-2\psi}\left[ e^{2\gamma}(x^2-y^2)\left( \frac{dx^2}{x^2-1} + \frac{dy^2}{1-y^2} \right) 
+ (x^2-1)(1-y^2) d\varphi^2\right] \ ,
\label{lel}
\ee
where the metric functions $\psi$ and $\gamma$ depend on the spatial coordinates $x$ and $y$, only, and $\sigma$ represents a non-zero real constant.
The corresponding vacuum field equations can be written as
\be
[(x^2-1)\psi_x]_x + [(1-y^2)\psi_y]_y =0\ ,\quad \psi_x = \frac{\partial\psi}{\partial x} \ ,
\label{eqpsixy}
\ee
\be
\gamma_x = \frac{1-y^2}{x^2-y^2}\left[ x(x^2-1)\psi_x^2 - x(1-y^2)\psi_y^2 -2y(x^2-1)\psi_x\psi_y\right]\ ,
\ee
\be
\gamma_y = \frac{1-y^2}{x^2-y^2}\left[ y(x^2-1)\psi_x^2 - y(1-y^2)\psi_y^2 +2x(1-y^2)\psi_x\psi_y\right]\ .
\ee
It can be seen that the function $\gamma$ can be calculated by quadratures once $\psi$ is known. If we demand that $\psi$ be asymptotically flat, i.e.,
\be 
\lim_{x\to\infty} \psi(x,y)=0 \ ,
\ee
it can be shown \cite{quev90} that using quadratures the asymptotically flat function $\gamma$ can be calculated as 
\be
\gamma= (x^2-1)\int_{-1}^{y}(x^2-y^2)^{-1}\left[ y(x^2-1)\psi_x^2 - y(1-y^2)\psi_y^2 +2x(1-y^2)\psi_x\psi_y\right] dy\ .
\ee

Suppose that a solution $\psi_0$ and $\gamma_0$ of this system is known. It is then easy to see that $\psi=\delta\psi_0$ and $\gamma = \delta^2 \gamma_0$
is also a solution for any constant $\delta$. This is the Zipoy-Voorhees transformation that can be used to generate new solutions. The simplest 
example is 
\be
\psi = \frac{\delta}{2}\ln\frac{x-1}{x+1}\ ,\quad \gamma = \frac{\delta^2}{2}\ln \frac{x^2-1}{x^2-y^2} \ ,
\label{zv}
\ee
which is generated from Schwarzschild solution ($\delta=1$). This metric is known in the literature as the $\delta-$metric to emphasize the fact 
that it is obtained by applying a Zipoy-Voorhees transformation with constant $\delta$.

A different representation can be obtained by using cylindrical coordinates that are defined as
\be
\rho = \sigma\sqrt{(1-y^2)(x^2-1)}\ ,\quad z = \sigma x y\ .
\ee
and in which the line element becomes
\be 
ds^2 = e^{2\psi} dt^2 - e^{-2\psi} \left[ e^{2\gamma} (d\rho^2 + dz^2) + \rho^2 d\varphi^2\right]\ .
\ee
Then, the vacuum field equations can be expressed as
\be
\psi_{\rho\rho} + \frac{1}{\rho}\psi_\rho + \psi_{zz} = 0\ ,
\ee
\be
\gamma_\rho = \rho( \psi_\rho ^2 - \psi_z^2) \ , \quad \gamma_z = 2 \rho \psi_\rho \psi_z\ .
\ee
In this representation, the Zipoy-Voorhees metric can be expressed in the Weyl form \cite{solutions}
\be
\psi = \sum_{n=0}^\infty \frac{a_n}{(\rho^2+z^2)^\frac{n+1}{2}} P_n({\cos\theta}) \ ,
\qquad \cos\theta = \frac{z}{\sqrt{\rho^2+z^2}} \ ,
\label{weylsol}
\ee
\be
\gamma = - \sum_{n,m=0}^\infty \frac{ a_n a_m (n+1)(m+1)}{(n+m+2)(\rho^2+z^2)^\frac{n+m+2}{2} }
\left(P_nP_m - P_{n+1}P_{m+1} \right) \ .
\ee
where $a_n$ $(n=0,1,...)$ are arbitrary constants, and $P_n(\cos\theta)$ represents the Legendre
polynomials of degree $n$. The Zipoy-Voorhees metric can be obtained by choosing the constants
$a_n$ in such a way that the infinite sum (\ref{weylsol}) converges to (\ref{zv})
in cylindric coordinates. A simpler  representation, however, is obtained in spherical coordinates
which are defined by means of the relationships
\be
\rho^2= (r^2-2\sigma r) \sin^2\theta \ , \quad z  = (r-\sigma) \cos\theta\ ,
\ee
so that the metric becomes
\be 
\label{zvspher}
ds^2 = \Delta^\delta dt^2 - \Delta^{1-\delta}\left[ \Sigma^{1-\delta^2} \Delta^{\delta^2-1}\left(\frac{dr^2}{\Delta} + r^2 d\theta^2 \right) + r^2\sin^2\theta d\varphi^2\right]\ ,
\ee
\be
\Delta = 1- \frac{2\sigma}{r}\ ,\quad \Sigma = 1- \frac{2\sigma}{r} + \frac{\sigma^2}{r^2}\sin^2\theta\ .
\ee
An analysis of the Newtonian limit of this metric shows that it corresponds to a thin rod source of constant density $\delta$, 
uniformly distributed along the $z-$axis from $z_1=-\sigma$ to $z_2=\sigma$. In the literature, usually a different constant $\gamma$ is
used instead of $\delta$, and, therefore, the Zipoy-Voorhees metric in the representation (\ref{zvspher}) is known as the Gamma-metric.

\section{The $q-$metric}
\label{sec:qmet}
If we start from the Schwarzschild solution and apply a Zipoy-Voorhees transformation with $\delta = 1+q$, we obtain the metric
\be
ds^2 = \left(1-\frac{2m}{r}\right)^{1+q} dt^2  
- \left(1-\frac{2m}{r}\right)^{-q}\left[ \left(1+\frac{m^2\sin^2\theta}{r^2-2mr}\right)^{-q(2+q)} \left(\frac{dr^2}{1-\frac{2m}{r}}+ r^2d\theta^2\right) + r^2 \sin^2\theta d\varphi^2\right].
\label{qmet}
\ee 
In \cite{quev11}, it was shown that this is the simplest generalization of the Schwarzschild solution that contains the additional parameter $q$, 
which describes the deformation of the mass distribution. In fact, this can be shown explicitly by calculating the invariant Geroch multipoles 
\cite{ger}. The 
lowest mass multipole moments $M_n$, $n=0,1,\ldots $ are given by
\be 
M_0= (1+q)m\ , \quad M_2 = -\frac{m^3}{3}q(1+q)(2+q)\ ,
\ee
whereas higher moments are proportional to $mq$ and can be 
completely rewritten in terms of $M_0$ and $M_2$. Accordingly, the arbitrary parameters $m$ and $q$ determine the mass and quadrupole 
which are the only independent multipole moments of the solution. In the limiting case $q=0$ only the monopole $M_0=m$ 
survives, as in the Schwarzschild spacetime. In the limit $m=0$, with $q\neq 0$, all moments vanish identically, implying that 
no mass distribution is present and the spacetime must be flat. The same is true in the limiting case $q\rightarrow -1$ which corresponds
to the Minkowski metric. 
Notice that all odd multipole moments are zero because the solution possesses an additional 
reflection symmetry with respect to the equatorial plane $\theta=\pi/2$.  

The deformation is described by the quadrupole moment $M_2$ which is positive for a prolate source 
and negative for an oblate source. This implies that the parameter $q$ can be either positive or negative. 
Since the total mass $M_0$ of the source must be positive, we must assume that $q>-1$ for positive values of $m$, and $q<-1$ for negative values of $m$.
We conclude that the above metric can be used to describe the exterior gravitational field of a static positive mass $M_0$ with a positive or negative
quadrupole moment $M_2$. The behavior of the mass moments depends on the explicit value of $q$. We will refer to the metric (\ref{qmet}) as
to the $q-$metric to the emphasize its physical significance as the simplest solution with an independent quadrupole moment.

The Kretschmann scalar
\be
K  = R_{\mu\nu\lambda\tau}R^{\mu\nu\lambda\tau} = \frac{16 m^2(1+q)^2}{r^{4(2+2q+q^2)}}\frac{ (r^2-2mr+m^2\sin^2\theta)^{2(2q+q^2)-1}}{(1-2m/r)^{2(q^2+q+1)}}L(r,\theta)\ ,
\label{kre}
\ee
\bea
L(r,\theta)= & & 3(r-2m-qm)^2(r^2-2mr+m^2\sin^2\theta) \nonumber\\
& & + m^2 q(2+q)\sin^2\theta[ m^2 q(2+q) + 3(r-m)(r-2m-qm)] \ ,
\eea
can be used to explore the singularities of the spacetime. We can see that only the cases $q=-1$ and $m=0$ are free of singularities. In fact, as noticed above, these cases correspond to a flat spacetime. A singularity exists at $r\rightarrow 0$ for any value of $q$ and $m$. In fact, for negative values of $m$
this is the only singular point of the spacetime which thus describes a naked singularity situated at the origin. In the range $2q(2+q)<1$ with $m>0$, 
there is singularity at those values of $r$ that satisfy the condition  $r^2-2mr+m^2\sin^2\theta=0$, i.e, these singularities are all situated inside a
sphere of radius $2m$. Finally, an additional singularity appears at the radius $r=2m$ 
which, according to the metric (\ref{qmet}), is also a horizon in the sense that the norm of the timelike Killing 
tensor vanishes at that radius. Outside the hypersurface $r=2m$ no additional horizon exists, indicating 
that the singularities situated at $r=2m$ and inside this sphere are naked. This result is in accordance with the black holes uniqueness theorems which 
establishes that the only compact object possessing an event horizon that covers the inner singularity is described by the Schwarschild solution. 
 
The position of the outer most singularity situated at $r_s=2m$ can be evaluated by using the expression for the invariant mass, i.e, $r_s=2M_0/(1+q)$.
In astrophysical compact objects, one expects that the quadrupole moment is small so that $q<<1$. Then the radius $r_s$ of the singular sphere is of the order of magnitude of the Schwarzschild radius $2M_0$ of a compact object of mass $M_0$, which is usually located well inside the matter distribution. It follows that 
it should be possible to ``eliminate" the naked singular sphere by finding the interior metric of an appropriate matter distribution that would fill completely 
the singular regions.

\section{The interior metric}
\label{sec:int}

It is very difficult to find physically reasonable solutions in general relativity, because the underlying differential equations 
are highly nonlinear with very strong couplings between the metric functions. In \cite{quev12}, a numerical solution was derived
for a particular choice of the interior static and axially symmetric  line element
\be
ds^2 = fdt^2 - \frac{e^{2k_0}}{f}\left(\frac{dr^2}{h} + d\theta^2\right) -\frac{\mu^2}{f}d\varphi^2\ ,
\ee
where 
\be
e^{2k_0} = (r^2-2mr+m^2\cos^2\theta)e^{2 k (r,\theta)}\ ,
\ee
and $f=f(r,\theta)$, $h=h(r)$, and $\mu=\mu(r,\theta)$.  To solve Einstein's equations with a perfect fluid source, the  
pressure and the energy must be functions of the coordinates $r$ and $\theta$. 
However, if we assume that 
$\rho=$ const, the complexity of the corresponding differential equations reduces drastically:
\be
p_r = - \frac{1}{2} (p+\rho) \frac{f_r}{f}\ , \quad p_\theta =  - \frac{1}{2} (p+\rho) \frac{f_\theta}{f}\ ,
\ee
\be
\mu_{rr} = -\frac{1}{2h} \left( 2 \mu_{\theta\theta} + h_{r} \mu_r - 32 \pi p\frac{\mu e^{2\gamma_0}}{f} \right) \ ,
\ee
\be
f_{rr} = \frac{f_r^2}{f} -\left(\frac{h_r}{2h} + \frac{\mu_r}{\mu}\right)f_r + \frac{f_\theta^2}{hf} -
\frac{\mu_\theta f_\theta}{\mu h} -\frac{f_{\theta\theta}}{h} + 8\pi \frac{(3p+\rho)e^{2\gamma_0}}{h}\ .
\ee
In addition, the function $k$ is determined by a set of two partial differential equations
which can be integrated by quadratures once $f$ and $\mu$ are known. The integrability condition of these partial
differential equations turns out to be satisfied identically by virtue of the remaining field equations.
It is then possible to perform a numerical integration by imposing
appropriate initial conditions. In particular, if we demand that the metric functions and the pressure are finite 
at the axis, it is possible to find a class of numerical solutions which can be matched with the exterior $q-$metric 
with a pressure that vanishes at the matching surface.

A different approach consists in postulating the interior line element and evaluating the energy-momentum tensor from 
the Einstein equations. This method was first proposed by Synge and has been applied very intensively to find approximate 
interior solutions \cite{bqr12,mala04}. To find the interior metric we proceed as follows. Consider the case of a slightly deformed mass. 
This means that the parameter $q$ can be considered as infinitesimal and this fact can be used to construct the interior 
metric functions. In fact, to the zeroth-order an interior line element can be obtained just by assuming that instead of
the constant $m$, the function $\mu(r)$ appears in the metric.
In the case of the $q-$metric, the functions entering the metric can be separated as
\be
\left(1-\frac{2m}{r}\right)\left(1-\frac{2m}{r}\right)^{c_1q+c_2q^2} \ ,
\ee
where $c_1$ and $c_2$ are constants. Then, to the first order in $q$, we can approximate this combination of functions as
\be
\left(1-\frac{2\mu}{r}\right)[1+c_1q\alpha(r)] \ .
\ee
Following this procedure, an appropriate interior line element for the $q-$metric (\ref{qmet}) can be expressed as
\be
ds^2 =  \left(1-\frac{2\mu}{r}\right)(1+q\alpha) dt^2 - (1+q\alpha+q\beta)\left(\frac{dr^2}{1-\frac{2\mu}{r}} + r^2d\theta^2\right)
-r^2\sin^2\theta (1-q\alpha) d\varphi^2\ ,
\label{int}
\ee
where 
$\mu=\mu(r)$, $\alpha=\alpha(r)$ and $\beta=\beta(r,\theta)$. 

Let us now consider the boundary conditions at the matching surface by comparing the above
interior metric (\ref{int}) with the $q-$metric to first order in $q$, i.e.,
\bea
ds^2= & & \left(1-\frac{2m}{r}\right)\left[ 1+ q\ln\left(1-\frac{2m}{r}\right)\right] dt^2 
- r^2 \left[ 1- q\ln\left(1-\frac{2m}{r}\right)\right] d\varphi^2 \nonumber\\
& -&  \left[ 1 + q  \ln\left(1-\frac{2m}{r}\right) - 2q \ln\left(1-\frac{2m}{r} + \frac{m^2}{r^2}\sin^2\theta\right)\right ] 
\left(\frac{dr^2}{1-\frac{2m}{r}} + r^2 d\theta^2\right) \ .
\label{qmatch}
\eea
A comparison of the metrics (\ref{int}) and (\ref{qmatch}) shows that they coincide at the matching  radius $r=r_m$, if the conditions
\be
\mu(r_m) = m \ ,\quad \alpha(r_m) = \ln \left(1-\frac{2m}{r_m} \right) \ ,
\quad \beta(r_m,\theta) = - 2 \ln\left(1-\frac{2m}{r_m} + \frac{m^2}{r_m^2}\sin^2\theta\right)\ ,
\ee
are satisfied.
Notice that we reach the desired matching by fixing only the radial coordinate as $r=r_m$, but it does not mean that the matching 
surface is a sphere. Indeed, the shape of matching surface is determined by the conditions $t=const$ and $r=r_m$ which, according to Eq.(\ref{qmatch}), 
determine a surface with explicit $\theta-$dependence.  

Finally, we calculate the Einstein tensor $G_\nu^\mu$ and find that the only non-diagonal component $G^r_\theta$ implies the equation
\be
\beta_\theta = \frac{r\cos\theta}{\sin\theta}\frac{(r-2\mu)(2\alpha_r+ \beta_r)}{r\mu_r - r +\mu}
\ee
which partially determines the function $\beta(r,\theta)$. Furthermore, the energy conditions $T^t_t\geq 0$ and $T^t_t - T^r_r\geq 0$ lead to
\be 
q\left[ \beta_{\theta\theta} + r\beta_{rr} (r-2\mu) - \beta_r (r\mu_r +\mu -r) + 4\mu_r (\alpha+\beta) - 2 r\alpha_r\right]\geq 4 \mu_r\ ,
\ee
\be
q\left[ \beta_{\theta\theta} + r\beta_{rr} (r-2\mu) + 4\beta \mu_r + 4\mu \alpha_r + \cot\theta \beta_\theta\right]\geq 0\ ,
\ee
respectively. A preliminary numerical analysis of these equations shows that it is possible to find solutions that satisfy the boundary conditions and the energy conditions simultaneously. In fact, the pressure and the energy density obtained in this way show a profile that is in accordance with the physical expectations.  
We conclude that by applying Synge's method it is possible to find physically reasonable interior solutions for the exterior $q-$metric. However,  it will be
necessary to further analyze the numerical solutions to find the ranges of boundary values of the main physical parameters that one can use to obtain physical 
configurations.

\section{Conclusions}
\label{sec:con}

In this work we discussed the Zipoy-Voorhees metric in different coordinate representations. 
We propose a different interpretation in terms of the quadrupole parameter $q$ and, therefore,
we designate it the $q-$metric. We found all the singularities of the underlying spacetime.
It was shown that only the Minkowski spacetime is free of curvature singularities, and that only   
the Schwarzschild spacetime possesses an event horizon that separates the inner singularity from 
the exterior spacetime. For all the remaining cases with non-vanishing quadrupole moment, it
was established that naked singularities are present inside a sphere with a radius which 
is of the same order or magnitude of the Schwarzschild radius for astrophysical compact objects.

We investigated the possibility of finding interior metrics that could be matched with the exterior
$q-$metric. In particular, we postulated a specific line element for the interior metric and
used Synge's method to derive the matter distribution. The matching conditions and the energy 
conditions were calculated explicitly in the case of a deformed source with a small quadrupole parameter.
It was shown that the resulting system of differential equations is compatible and that 
particular solutions can be calculated by using numerical methods.

The resulting system of differential equations for the functions of the interior metric indicates 
that one can try to find analytical solutions, at least in the case of a slightly deformed mass
distribution. To do this, it will be necessary to investigate in detail the mathematical 
properties of the  differential equations. This is a task for future investigations.

Moreover, we expect to apply the same method in the case of rotating sources. The 
rotating $q-$metric was derived in \cite{quev90}, but no attempts have been made to 
investigate its physical properties and the possibility of matching it with a
suitable interior metric. This problems will be the subject of future research.

\section*{Acknowledgments}
This work was supported in part by Conacyt, Grant No. 166391.


\begin{thebibliography}{99}

\bibitem{zip66} D. M. Zipoy, 
J. Math. Phys. {\bf 7} (1966) 1137.

\bibitem{voor70} B. Voorhees, 
Phys. Rev. D {\bf 2} (1970) 2119.
 
\bibitem{solutions}
H. Stephani, D. Kramer, M. A. H. MacCallum, C. Hoenselaers, and E. Herlt, 
{\it Exact Solutions of Einstein's Field Equations} (Cambridge University Press, Cambridge, 2003).

\bibitem{zv1} D. Papadopoulos, B. Stewart, L. Witten, Phys. Rev. D {\bf 24} (1981) 320. 
 
\bibitem{zv2} L. Herrera and J. L. Hernandez-Pastora, J. Math. Phys. {\bf 41} (2000) 7544. 

\bibitem{zv3} L. Herrera, G. Magli and D. Malafarina, Gen. Rel. Grav. {\bf 37} (2005) 1371.

\bibitem{zv4} N. Dadhich and G. Date, (2000),  arXvi:gr-qc/0012093 


\bibitem{zv5} H. Kodama and W. Hikida, Class. Quantum Grav. {\bf 20} (2003) 5121.

\bibitem{quev11} H. Quevedo, Int. J. Mod. Phys. {\bf 20} (2011) 1779.

\bibitem{synge} J. L. Synge, Relativity: The General Theory (North-Holland, Amsterdam, 1960).

\bibitem{quev90} H. Quevedo, 
Forts. Physik {\bf 38} (1990) 733.


\bibitem{ger} R. Geroch, 
{J. Math. Phys.} {\bf 11} (1970) 1955;
{J. Math. Phys.} {\bf 11} (1970) 2580.


\bibitem{quev12} H. Quevedo, {\it Multipolar Solutions}, in Proceedings of the XIV Brazilian School of Cosmology and Gravitation, (2012);
arXiv:1201.1608
 
 
\bibitem{bqr12} K. Boshkayev, H. Quevedo, and R. Ruffini, Phys. Rev. D {\bf 86}:064403 (2012).

\bibitem{mala04} D. Malafarina, Conf. Proc. C0405132 (2004) 273.
 
 

\end{thebibliography}
\end{document}